# MODULATION CLASSIFICATION THROUGH DEEP LEARNING USING RESOLUTION TRANSFORMED SPECTROGRAMS

**Muhammad Waqas[1], Muhammad Ashraf[1], Muhammad Zakwan[1]**
[1]Department of Avionics Engineering, Air University, Islamabad
Corresponding author: Muhammad Waqas (e-mail: *212110@students.au.edu.pk*)
Corresponding author: Muhammad Zakwan (e-mail: *mzakwan@mail.au.edu.pk*).

**ABSTRACT** Modulation classification is an essential step of signal processing and has been regularly applied in the field of tele-communication. Since variations of frequency with respect to time remains a vital distinction among radio signals having different modulation formats, these variations can be used for feature extraction by converting 1-D radio signals into frequency domain. In this paper, we propose a scheme for Automatic Modulation Classification (AMC) using modern architectures of Convolutional Neural Networks (CNN), through generating spectrum images of eleven different modulation types. Additionally, we perform resolution transformation of spectrograms that results up to 99.61% of computational load reduction and 8x faster conversion from the received I/Q data. This proposed AMC is implemented on CPU and GPU, to recognize digital as well as analogue signal modulation schemes on signals. The performance is evaluated on existing CNN models including SqueezeNet, Resnet-50, InceptionResnet-V2, Inception-V3, VGG-16 and Densenet-201. Best results of 91.2% are achieved in presence of AWGN and other noise impairments in the signals, stating that the transformed spectrogram-based AMC has good classification accuracy as the spectral features are highly discriminant, and CNN based models have capability to extract these high-dimensional features. The spectrograms were created under different SNRs ranging from 5 to 30db with a step size of 5db to observe the experimental results at various SNR levels. The proposed methodology is efficient to be applied in wireless communication networks for real-time applications.

## I. INTRODUCTION

With rapid development of wireless communication systems coupled with limited availability of channels, these wireless systems have been forced to use intensively dense connected networks with aggressive spectrum utilization, which has led to a number of undesirable issues such co-channel interference and signal distortion over propagation channels. In these conditions, providing the regulatory bodies with the ability to classify the modulation type of the signal received can aid in user identification and occasionally even enable information retrieval from non-encrypted communications. The usage of AMC is vital to tele-communication systems, especially in the 5G and IoT age, as it is required for demodulating signals at the receivers' end [1]. While cooperative and non-cooperative conditions can be used to classify modulation scheme of received signal; modulation classification or recognition functions as an essential step between signal detection and demodulation in non-cooperative conditions [2].

In a highly congested Electromagnetic (EM) environment, current spectrum monitoring techniques employed in the fields of radar and communication are capable of gathering enormous RF data in wideband using their cutting-edge technologies. The most recent methods, however, have difficulty achieving the necessary levels of accuracy at a computational cost that would render their implementation on low-cost platforms. These methods employ such data to identify emitter kinds and detect communication schemes. Due to its notable successes in a number of areas, including speech recognition, sentiment analysis, and computer vision; Machine Learning (ML) is gaining attention [3]. Most significantly, Deep Learning (DL) has demonstrated enormous promise compared to previous methods of ML in terms of data comprehension because it automatically improves representation of complicated, high-dimensional data without manually extracting the features [4].

In the civilian field, AMC is mainly used for interference identification, signal confirmation, and interference confirmation for EM spectrum management. Besides, it is an important tool for amateur satellite receiving stations. In the military field, it is used for identification of threat type, providing Communication Intelligence (COMINT), spoofing in enemy's radio, triangulation or localizing the transmitter, and various Electronic Countermeasures (ECM).

Recently, AMC has gained more attention than ever for developing modern systems like Software Defined Radio (SDR).

The fast advancement of machine learning has attained the attention due to its substantial achievements in various applications; however, compared in terms of data comprehension, DL has shown immense potential as it automatically obtains features of complex and high-dimensional data without manually extracting the features [4]. The processing time-line of DL-based classification is depicted in figure 1.

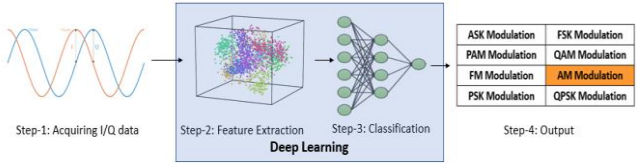

**FIGURE 1.** Processing time-line of ML and DL-based AMC application.

This research was conducted to understand the challenges of DL based AMC and to propose a solution, it has following key contributions:

1)  This research provides a survey of all existing DL based AMC researches investigating advanced structures and modern designs of deep networks for various types of data pertaining to sequential radio signals including I/Q samples, spectrum images, time-frequency graphs, eye diagrams and constellation images used for AMC purposes.

2)  It compares the results of latest works on the basis of DL models-utilized vis-à-vis accuracy attained, while classifying different modulation formats.

3)  It then proposes a solution by generating spectrograms from I/Q samples of signal data and transforming their resolution for speedy conversion and lower computational requirements, that can be used without filtration, in real-time modulation classification applications.

4)  It validates the recognition accuracy of these transformed spectrograms through implementation on existing DL models and highlights research challenges as well as potential way forwards in the field of DL based AMC.

The paper is structured as follows: Section II elaborates revolutionary approaches for DL based AMC, Section III cites the literature review, Section IV cites the problem statement, Section V states the proposed methodology, Section VI cites the experimental setup, Section VII provides the simulation results while Section VIII provides the conclusion and potential future directions to this work.

## II. REVOLUTIONARY APPROACHES FOR AMC

### A FEATURE-BASED CLASSIFICATION

It is feasible to divide the algorithms used to solve the AMC problem into two groups: likelihood-based (LB) and feature-based (FB). The likelihood ratio is a measure of how much more likely one hypothesis is compared to another, given the observed data. It is calculated by dividing the likelihood of the data under one hypothesis by the likelihood of the data under another hypothesis. The resulting value is then compared to a threshold value to determine which hypothesis is more likely [5]. Though, the LB algorithms have optimal performance, their limited applicability is caused by their high processing requirements [6]. Contrary to it, the FB algorithms make use of the unique and distinctive features to classify modulation type. The organogram of feature-based modulation identification elaborating these types is given in figure 2.They are more popular in AMC because they have shown potential for achieving better results with a much lower complexity [5]. Summary of comparison of these two traditional methods for AMC is given in Table I.

TABLE I
COMPARISON OF LIKELIHOOD VS. FEATURE BASED CLASSIFICATION

| Properties | Likelihood-based classification | Feature-based classification |
|---|---|---|
| Computation requirement | Large | Small |
| Identification matrix | Simple | Complex |
| Theoretical base | Available | Limited |
| Effect of noise | Less effect | More effect |
| Classification accuracy | Satisfactory | Good |

### B DL vs ML ALGORITHMS FOR CLASSIFICATION

DL is a representation learning algorithm that is built on massive amounts of ML data. By using an unsupervised or semi-supervised feature learning algorithm, DL can extract the features of the data. Traditional ML techniques, on the other hand, require the manual extraction of features, thereby adding the workload. The DL-based AMC algorithms focus on feeding a huge dataset into the model, as effectively as possible. Additionally, the developed network learns and extracts the signal's properties from the provided dataset before classifying according to the designed algorithm [2]. It has been observed that the DL-based AMC methods have many advantages, thereby enabling many scholars to design different DL networks for recognition [7]. The comparison between ML-based AMC approaches and DL-based AMC is listed in table II.

TABLE II
COMPARISON OF ML AND DL BASED CLASSIFICATION

| Properties | ML-based classification | DL-based classification |
|---|---|---|
| Expert knowledge | Required | Not required |
| Feature Extraction | Manual / Less | Automatic / Enhance |
| Stages of model | Feature extraction & classification | End-to-end system (single stage) |
| Classification Accuracy | Inferior | Superior |
| Computational Complexity | Simple | Complex |
| Training Data Requirements | Small amount | Large amount |

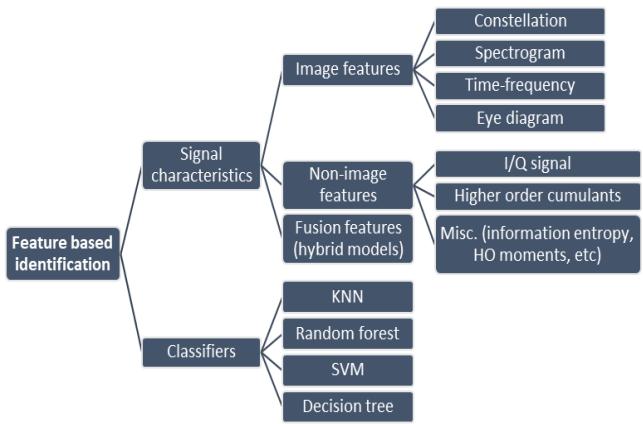

**Figure 2.** Organogram / tree of feature-based AMC.

*C DATA DEPENDENCE AND AVAILABILITY*

One of the most important issues with DL is data dependence since, in comparison to conventional ML techniques, it is more dependent on the training data and consequently needs a lot of data to comprehend the latent patterns inside. An intriguing finding is that there is practically a linear relationship between the size of the required amount of data and the model's scale [4]. Different open-source modulation signal datasets were examined for this research including *RadioML2016.10a, RadioML2016.10b, RadioML2016.04c, RadioML2018.10A* and *HisarMod2019.1*. It is important to note that several DL nets, like as CNN, are effective for processing images because they can quickly extract feature information from images using convolution filters [7]. In order to determine the type of modulated signal, researchers typically convert the signal into two-dimensional images like eye diagrams and constellation images, followed by use of convolutional layers to extract the signal's features from the image, and then fully connected layers to perform the classification. However, many academics utilize CNN to directly extract characteristics from signals due to the utility of the application being developed [7], [8].

## III. LITERATURE REVIEW

*A TYPES OF DL MODELS USED FOR AMC*

Recent researchers have focused on multiple ways of extracting features for AMC; these extraction can be *spectral* features [8]–[12], *constellation* features [13]–[18], *statistical (I/Q)-domain* features [19]–[25], or *time-frequency* features [26]–[30] with acceptable level of accuracy. *Different categories of DL networks* has been utilized in this domain; so, for the ease of understanding and comparison, we can segregate the reviewed works on DL-based AMC into Convolutional Neural Networks **(CNNs)** [10]–[14], [26], [30]–[35], Recurrent Neural Networks **(RNNs)** [9], [16], [24] [22], [36], [37] and **hybrid models** [8], [14], [18]–[20], [23]–[25], [29]. AMC's model-wise categorization summary is listed in table III.

TABLE III
MODEL WISE CATEGORIZATION OF REVIEWED AMC RESEARCHES

| Categories of DL networks | CNNs | RNNs / LSTM | Hybrid models |
|---|---|---|---|
| References of researches | [10]–[14], [26], [30]–[35] | [9] [16] [22] [24] [26] [36] [37] [38] | [8] [18] [14] [19] [20] [23], [24] [25] [29] |
| Researches reviewed | 13 | 08 | 09 |

The classification performances of these cited researches vary with the data set representation or the type of dataset used. This suggests that investigating available data representations is necessary to propose a suitable solution for AMC. This, in addition to identify the gaps in recent researches of AMC in DL domain, encouraged us to survey and compare reviewed works on the basis of type of data / input used vis-à-vis the model architecture and the attained results.

*B UTILIZING SPECTROGRAM AS INPUT*

A spectrogram is considered as a very accurate and finely detailed representation of signal data. A typical spectrogram is an image with a time axis and a frequency axis, and the colour of each point corresponds to the amplitude of the signal there. As a result, a spectrogram can show the signal's strength in relation to each of its frequency components. Because stationary noise has a lower value at nonzero cyclic frequency and a cyclic spectrum has high anti-noise performance, it is frequently employed to evaluate signals in surroundings that are noisy and cluttered [8]–[12]. We may convert 1-D radio signals into spectrogram images using STFT because the frequency variation with time is the key differentiator among radio signals with various modulation types. Many recent researchers, that will be discussed ahead, have therefore utilized spectrogram images as input to the neural networks.

In [10], researchers used the cyclic spectrum and an innovative CNN model based on spectral features. The results indicate that CNN has good performance in extracting high-dimensional features, however *these results were not compared / validated against any other architecture*. In [11], researchers presented a time-frequency analysis of modulated radio signals and designed a spectrum based convolutional neural network (SCNN) framework for AMC purpose classifying 11 modulation types. In [8], the cyclic spectrum was fed into CNN and the parallel network of GRU for modulation classification; thereby combining the recognition characteristics of the two networks (CNN and GRU). [12] demonstrated the modulation recognition by CNN through image data enhancement, whereby classifying 07 modulation schemes of RML2016.10a data excluding 16QAM, 64QAM and QPSK schemes from the dataset. For classification of different radar waveforms, a phase-based method applying

Short Time Fourier Transform (STFT) and Bidirectional Long Short Term Memory (BiLSTM) was presented in [9]. Table IV shows the comparison of these discussed spectrogram-based researches.

TABLE IV
SPECTROGRAM BASED APPROACHES OF AMC USING DL METHODS

| Research Detail | Ref | Model Used | Mod. Type | Accuracy Achieved |
|---|---|---|---|---|
| Utilizing CNN algorithm | [10] | 5-layer CNN | **08** | 90.4% at -2db 75% at-8db |
| Spectrum Analysis and CNN | [11] | 4-layer CNN | **11** | 80% at 0db <60% at-5db |
| AMC Based on CNN & GRU | [8] | CNN and GRU | **08** | 78% at –6dB 96% at 0 dB 100% at 4 dB |
| Method of Complex Modulation Signal Based on CNN | [12] | CNNs AlexNet VGG16 VGG19 Resnet18 | **07** | At 5 dB: 82% 86% 87% 91% |
| Using Phase-Based STFT and BiLSTM | [9] | BiLSTM | **06 Radar Waveforms** | 83% at 0 dB |

*C UTILIZING CONSTELLATION IMAGES AS INPUT*

By converting signal samples into scattering points on a complex two-dimensional plane, a constellation image can be created as a popular 2-D depiction of a modulated signal. It uses the amplitude and phase information of its sample points to represent a modulated signal. However, the sample points from the modulated signal do not converge in a single point in the Constellation Diagram in the presence of noise and off-set, causing an error in its amplitude and phase information.

In [18], researchers proposed an Auxiliary Classifier Generative Adversarial Networks (ACGANs) using an existing CNN model (AlexNet) as classifier to enlightened the effect of reducing the original dataset and then compensating it by using the generated images from ACGAN, to attain good classification results. In [16], researchers proposed a simpler graphic constellation projection (GCP) for classification of *only four types of modulation schemes*. [14] used two CNN-based DL models (AlexNet and GoogLeNet) and demonstrated the effect of image resolution on the classification results; however, *classification of MQAM signals remained degraded (<80%) even during only eight types of available modulation schemes*. In [17], a unified mesh model for the constellation diagrams of only M-QAM signals to attain Good classification results with the carrier perfectly synchronized; while with the *phase offset of π/16 and π /8, the accuracy achieved was mere 86% at 5 db*. Moreover, the model could *only classify amongst four M-QAM types*. In [15], researchers propose an integrated architecture called CLDNN combining CNN, LSTM and DNN networks; to demonstrate the results better than individual models; thereby supporting the claim that *combination networks may perform better in modulation classification*. [13] calculated the rotation degree of constellation image, while attaining accuracy of 97% at 15 dB SNR. Table V enlists the comparison summary of constellation-based AMC approaches.

TABLE V
CONSTELLATION BASED APPROACHES OF AMC USING DL METHODS

| Research Details | Ref | Model used | Mod. Types | Accuracy Achieved |
|---|---|---|---|---|
| AMC with Data Augmentation using GAN in Cognitive Radio Networks | [18] | AlexNet with GAN to generate data | **08** | 78% at –6dB 95% at -2 dB 99% at 0 dB |
| Graphic Constellations and DBN based AMC | [16] | DBN | **04** | 95% at 0db 100% at 3db |
| AMC Based on Signal Constellation and DL | [14] | AlexNet and GoogLeNet | **08** | **At 4db:** 16QAM=74% 32QAM=93% 64QAM=68% Rest all=100% |
| Graph-Based Constellation Analysis for M-ary QAM | [17] | Mesh Model (for M-QAM) | **07 (M-QAM)** | 85% at 0db 100% at 8db **With phase offset:** 85% at 5db 100% at 14db |
| AMC for satellite communication | [15] | CLDNN (CNN-LSTM-DNN) | **11** | 89% at 0db 90% at 6db |
| Artificial Feature Engineering DNN Modulation Identifier | [13] | CNN | **11** | 97% at 15 db |

*D UTILIZING IQ SAMPLE AS INPUT*

Two factors in particular make the use of an I/Q sample possible: Firstly, the features derived from I channel, Q channel, and I/Q multichannel data will be complimentary because the presence of unbalance between amplitude and phase impairs the orthogonal link between I and Q channel, causing differences between these two channels; secondly, the formats of these three-input data are consistent, and dividing the multi-channel data makes it simple to obtain independent I/Q channel inputs.

A hybrid model, Multi-Channel Convolutional Long Short-Term Deep Neural Network (MCLDNN) was proposed in [24] that integrates CNNs, LSTMs, and DNNs to fully extract features for classification. [23] combined the advantages of the CNN, and LSTM; however, at the expanse *of more time-consumption*. In [25], researchers first applied a Deep

Complex Network (DCN) to extract the features BiLSTM layers to build a memory model according to the extracted features. The resultant *accuracy on the 8PSK, 16QAM (error rate of 0.42) and 64QAM (error rate of 0.32) types was still not satisfactory* for practical applications. Moreover, the DCN-BiLSTM network showed *slow training speed*, and the method worked satisfactorily only for signals having limited frequency-offset and sampling frequency-offset. The method in [22] made use of an LSTM-based neural network architecture and was easily implemented on a low-cost computer platform. Compared to training using IQ components, training with L2-normalized amplitude and phase improved classification accuracy while using a great deal less computational power. In [20], researchers presented an ensemble of DL based modulation signal classification (EDL-MSC) models for acoustic underwater-communication; however, it can *only classify five types of modulation schemes* namely 8PSK, BPSK, 16QAM, 64QAM, and QPSK. In [19], researchers incorporated an attention mechanism for a hybrid model LSTM DNN (ASCLDNN) model for better computation time and accuracy. [38] exhibited a deep learning classifier that uses LSTM network classification and sub-Nyquist sampling AMC. Four datasets with various time-varying, 2-tap Rician fading channel models, route delays, and a Doppler shift were used to validate the performance. [37] concluded that *CNN is less efficient while extracting features from time-series signals*. Additionally, no significant improvement in results was achieved even after the extraction of *amplitude and phase information of I/Q signal, as the highest accuracy of CNN-AP was shown as 83.4%.* [36] showed and proposed RNN-based classifier, robust to the uncertain noise conditions and signal distortion for *four modulation schemes*. [32] demonstrated a cost-efficient CNN (MCNet) to classify 24 types of schemes. However, *higher schemes like 128APSK, attained poor accuracy* of 51.40% at +10 dB SNR, as got confused with 64APSK and 128QAM. *256QAM* was reported as the worst classified signal, and suffered confusion with 128APSK and 128QAM.

TABLE VI
I/Q SIGNAL-BASED APPROACHES COMPARISON WITH RESULTS

| Research Details | Ref | Model used | Mod. Type Classified | Accuracy Achieved |
|---|---|---|---|---|
| Spatiotemporal Multi-Channel Learning Framework | [24] | MLCDNN (CNN+ LSTM) | RML 2016.10a (11 types) | 89% at 0db 92% at 12dB |
| CNN-LSTM Based Dual-Stream Structure | [23] | CNN-LSTM | RML 2016.10a | 86% at 0db 90% at 4dB |
| DCN-BiLSTM Network | [25] | DCN-LSTM | RML 2016.10a | 90% at 4dB |
| Real-Time Radio Technology and AMC via an LSTM Auto-Encoder | [22] | LSTM | RML 2016.10A | 87% at 0db 91% at 4dB |
| | | | RML 2018.01A | 72% at 0db 97% at 4dB 99% at 12db |
| Ensemble of DL Enabled AMC | [20] | LSTM-GRU-SSAE | 05 | 87% at 0db 97% at 6dB |
| AMC Based on CLDNN | [19] | LSTM-CNN | RML 2016.10A | 87% at 0db 70% at -5dB 92% at 6dB |
| LSTM Guided AMC for Sub-Nyquist Rate Wideband Spectrum | [38] | LSTM (RNN) | 07 | 80% at 0 dB |
| LSTM With Random Erasing and Attention | [37] | LSTM | RML 2016.10a | At 4 dB: 83%- 1 layer 91%-2 layer 92%-1 layer with att'n |
| DNN under Uncertain Noise | [36] | RNN (LSTM) | 04 | 90% at 8 dB |
| MCNet: An Efficient CNN Architecture | [32] | CNN | 24 | <60% at 0 dB 85% at 10 dB |

*E UTILIZING TIME AND FREQUENCY SERIES IMAGES AS INPUT*

The features of the signal can be lost if only its temporal and frequency domain properties are examined. In order to thoroughly examine the pertinent components of the signal, researchers perform time-frequency analysis on the signal. This involves, for example, applying wavelet transform or other methods to convert the signal to a time-frequency map and extracting the features from these images.

In [29], the researchers performed time–frequency conversion on the LPI radar signal followed by dimensionality reduction by PCA (Principal Components Analysis) algorithm, to achieve a remarkable accuracy of 97% at 0 dB. In [30], researchers introduced the time-frequency analysis method on eight radar waveforms, and the TFI pre-processing methods, so as to facilitate the data processing of subsequent neural networks. In [33], a novel framework for AMC named ResNeXt network {which is an upgraded version of the residual network (ResNet)} was utilized, along with adaptive attention mechanism modules ; Due to the signals' short observation windows in RadioML2016 and silent periods with only carrier tones, the results indicated that AM-DSB and WBFM could not be differentiated. Moreover, due of their comparable time-frequency images and overlapped constellation points, QAM16 and QAM64 became indistinct.. [26] demonstrated a classification method combining time–frequency analysis and a DNN to identify radar modulation signals; however, the model *could only classify 07 types* of radar modulations.

With comparison given in table VII, the recognition rate of time-frequency based-models reaches good accuracy. However, these images have been mostly used against *signals of limited modulation schemes (associated with radar waveform)* and render degraded results when exposed to other schemes. Therefore, we need to pursue further optimization in this area in order to apply this strategy to real communication networks.

TABLE VII
TIME-FREQUENCY DIAGRAM-BASED APPROACHES COMPARISON WITH RESULT

| Research Details | Ref | Model used | Mod. Types | Accuracy Achieved |
|---|---|---|---|---|
| Radar Waveform Recognition | [29] | CNN-TPOT | 12 radar waveforms | 97% at 0db 99% at 6db |
| Radar Signal Intra-Pulse Mod Recognition | [30] | CNN | 08 radar waveforms | 94% at -6db |
| Adaptive Attention Mechanism | [33] | ResNext (upgraded version of ResNet) | RML 2016.10B and 2018.01A (10 types) | 90% at 0 db 74% at -4db |
| Radar Signal Modulation Recognition | [26] | Channel-separable ResNet | 07 radar waveforms | 88% at -13db 98% at -4db |

*F UTILIZING EYE DIAGRAM AS INPUT*

On the oscilloscope, a waveform is shown by a collection of digital signals. varied modulation signals have varied eye diagram properties, therefore it is possible to convert a signal into an eye diagram and then use a DL model to extract features from these images to accomplish modulation classification [39].

In [34], researchers presented results for Eye diagram based modulation recognition utilizing CNN along with concatenation of I and Q-eye diagram for classification of eight modulation schemes; however, with frequency offset (of ±0.01), reduced accuracy was observed at 6db SNR, while results for 16QAM were further degraded. *At 0 db, except BPSK and OQPSK, all modulation schemes had less than 75% results*. In [35], an eye-diagram analyser is proposed to implement both modulation format recognition (MFR) and optical signal-to-noise rate (OSNR) estimation by using a CNN-based DL method. However, the *results obtained were at high SNR (10-25dB) with only four modulation type studied.* A multi-input CNNs (ResNet) model was suggested by [31] to extract and map the features from several dimensions. The algorithm was able to perform well in low SNR conditions, although 64QAM signal performance was worse than other signals (30% at 0dB). Recognition accuracy of 100% at 0 dB was reached for BPSK and OQPSK signals due to their discriminative visual characteristics from other modulated signals in the eye diagram and the vector diagram. Comparison summary is given in table VIII.

TABLE VIII
EYE DIAGRAM-BASED APPROACHES COMPARISON WITH RESULTS

| Research Details | Ref | Model used | Mod. Types | Accuracy Achieved |
|---|---|---|---|---|
| Modulation recognition based on IQ-eye diagrams and DL | [34] | CNN | 08 | 90% at 6dB (except 16 QAM) <75% at 0dB |
| AMC and OSNR Estimation Using CNN-Based DL | [35] | CNN | 04 | 100% (10-25dB) |
| A DL Framework for Signal Detection and AMC | [31] | ResNet | 08 | 80% at 0 dB (30% for 64QAM) |

## IV. PROBLEM IDENTIFICATION

*A ANALYSIS OF EXISTING APPROACHES*

As discussed in the above section, recent researchers have explored the application of DL algorithms for AMC, and few have achieved quite acceptable results. To improve the classification results, one intuitive idea may be to *use more than one type of data sources*, such as spectrograms, I/Q samples, and eye-diagrams [31] [42, p. 50] [41]. It makes sense because they give new categorization bases for AMC models and provide various viewpoints of the modulation information for a signal. However, in practical applications, getting multiple signal forms is a challenging operation, and the algorithms' performance may be impacted by varying the input forms. Similarly, *time frequency diagram* and *eye diagram* have not been a viable approach when classifying more modulation schemes utilizing DL models. Forgone in view, challenges in this field found after literature review and comparative study of contemporary approaches lead to the following research gaps:

1) It has been observed that modulation classification results are *degraded severely by decreasing the SNR*, especially in the cases where identification types are 11 or more (with RML2016.10 dataset or upgraded one). Models giving good accuracy at low SNRs have only been trained and evaluated on lesser modulation types [8], [10], [11], [16], [18] [26], [29] [35].

2) Using *hybrid-models* for modulation classification helps in increasing the classification accuracy, however they suffer from computational complexity and slow training [24], [29], thereby limiting the utility of these models.

3) Utilizing *constellation diagram* as input to deep learning models can give accurate results in many cases, however classification on the basis of constellation generates confusion while classifying among M-QAM (16QAM, 64QAM, 128QAM) modulation types [14]. Also, any phase off-set ends in drastic reduction in results accuracy [17]. In light of this, this research work focuses on implementation of a DL based AMC that can classify diverse modulation schemes utilizing spectrogram images as input.

## V. PROPOSED METHODOLOGY

We investigated innovative AMC techniques where different deep models are developed as classifiers to not only mitigate the drawbacks of conventional AMC techniques but also to increase performance in terms of accuracy and complexity. Compared with conventional ML models, DL offers an advantages of automatic feature extraction and high learning capacity [20], [21]; thus increasing the classification accuracy of higher-order modulation formats under a synthetic channel deterioration [3]. As shown in figure 3, these inputs however need to undergo essential pre-processing before feeding into any DL network. The goal of data pre-processing is to create potentially useful wireless data representations by analysing and manipulating the collected spectrum data.

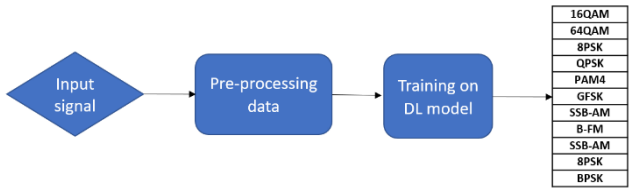

**Figure 3.** Overall processing flow of DL based AMC

Due to the capability of processing high-dimensional unstructured data, CNNs have proven their *suitability for images as inputs*, although they may be utilized for other applications with text or signals. Therefore, in the proposed model we will be focusing to exploit this capability of DL models and feed them with various forms of image input e.g. spectrogram images.

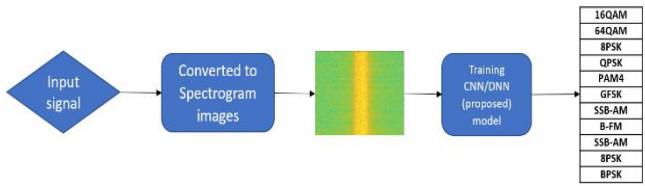

**Figure 4.** Overall processing flow of our proposed spectrogram-based AMC approach

*A UTILIZING TRANSFER LEARNING*
The purpose of transfer learning, a crucial domain in machine learning, is to transfer knowledge from one model to another by using similarities in data, tasks, or models. In contrast to conventional machine learning, it is independent of requirements for data annotations and can obey a variety of distributions. Additionally, it is possible to train new models and perform new tasks using the same features and weights that were collected from the training another model [42]. Transfer learning can be divided into four categories: sample-based, feature-based, model-based and relationship-based ones [2]; while utilized feature-based method only. Transfer learning has several benefits for DL-based AMC, including:

1)      IMPROVED CLASSIFICATION ACCURACY
Transfer learning can improve the classification accuracy of DL-based AMC by leveraging pre-trained models or layers that have already learned relevant features. [43] Used Transfer learning techniques on a CNN that was pre-trained on low-level features utilizing ImageNet dataset.
2)      REDUCED TRAINING TIME
Using pre-trained models or layers can reduce the amount of training time required for AMC models, as the network has already learned relevant features.
3)      ROBUSTNESS TO SIGNAL VARIABILITY
Transfer learning can improve the robustness of AMC models to signal variability by leveraging pre-trained models or layers that have learned features under various channel conditions.
In this research, applied transfer learning to different CNN models; followed by the fine tuning, and the transfer adaptation method. Cognizant of the fact that in any neural network structures, improvement in training can be attained by increasing the network depth; however, we avoided this method for improvement because increasing layers may cause issues like overfitting, high-training time, gradient explosion and so on. Aforesaid in view, we aim to extract features making use of less computational power and attaining acceptable results.

*B DATA SET SELECTION*
Real systems have a multitude of effects on the broadcast signal that make recovery and representation difficult since, in contrast, channel effects are not deterministic and not totally invertible in a communications system. With objective to select a realistic dataset for the proposed model training and to mimic a real data, the selected data has to be distorted due to following factors:
1)      THERMAL NOISE
Results in relatively flat white *Gaussian noise* at the receiver which forms a noise floor or sensitivity level.
2)      OSCILLATOR DRIFT
Due to temp / semiconductor physics differing at the transmitter and receiver result in symbol *timing offset, sample rate offset, carrier frequency offset and phase difference*.
3)      MULTI-PATH FADING
Real channels undergo random filtering based on the arriving modes of the transmitted signal at the receiver with varying amplitude, phase, Doppler, and delay.

Keeping in view the above factors, we selected RML2016.10.a dataset, that was proposed and published by Timothy J. O'Shea, Johnathan Corgan, and T. Charles Clancy in 2016 [44]. It uses well-defined transmit parameters to generate radio communications signals that are identical to real-world signals in terms of modulation, pulse shaping, communicated data, and other well-defined transmit characteristics. It contains 11 modulation types, encompassing 8 digital - BPSK, QPSK, 8PSK, 16QAM, 64QAM, BFSK, CPFSK, PAM4 and 3 analog schemes - WB-FM, AM-SSB, AM-DSB. This synthetic signal set is passed through harsh channel models which introduce unknown scale, translation, dilation, and impulsive noise onto the signal [45].

*C K-FOLD VALIDATION TO AVOID OVERFITTING*
The efficacy of categorization systems is frequently evaluated using the K-fold cross validation technique. A data collection is first arbitrarily divided into k folds with roughly equal numbers of examples. Each fold then tests the model created from the other k-1 folds. Given that the partition is random, the accuracy estimates' variance for statistical inference may be considerable. Based on the experimental findings on twenty data sets, it was demonstrated in [46] that k-fold cross validation be used to produce accurate accuracy estimates. [47] compared and analysed breast cancer classifications with different DL algorithms using k–Fold Cross Validation (KCV) technique. [48] and [49] proposed CNNs-based models for learning a shared representation among all the spectrograms from different classification tasks. [50] Applied ten-fold cross-validation strategy in order to train and test the models utilizing multi-channel spectrograms.

In our experiment, 5-fold cross-validation was used, and the dataset was divided into 5 folds with an equal number of samples in each fold. Each training cycle employed four folds to train the model, and the fifth fold was used as a validation dataset to fine-tune the model and assess training effectiveness. To make sure that every fold had been used to assess the model's performance, the procedure was performed five times.

| Fold 1 | Fold 2 | Fold 3 | Fold 4 | Fold 5 | Training cycles |
|---|---|---|---|---|---|
| Test | Train | Train | Train | Train | Iteration 1 |
| Train | Test | Train | Train | Train | Iteration 2 |
| Train | Train | Test | Train | Train | Iteration 3 |
| Train | Train | Train | Test | Train | Iteration 4 |
| Train | Train | Train | Train | Test | Iteration 5 |

**Figure 5.** Training cycles of 5-fold validation model.

## VI. EXPERIMENTAL SETUP

Subsequently, we aim to propose an algorithm, which provides a good recognition accuracy of digital signal modulation by combining transferring adaptation with already existing net in an efficient manner. Especially at different SNRs, we focused on the recognition rate of the RML2016.10A signal modulation modes. To accomplish the aforementioned methodology, cycle depicted in figure 6 was devised and followed.

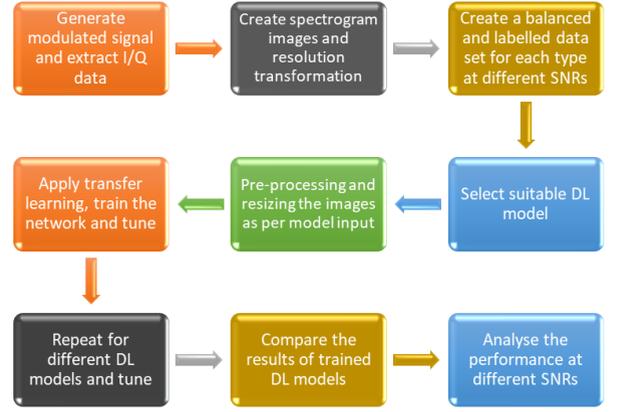

**Figure 6.** Experimental steps for DL based AMC model design using raw spectrograms.

*A GENERATING SPECTROGRAM IMAGES*
We initiated the task by finding the ideal parametric settings of spectrogram for feature learning through DL models. The ideal parameters for a spectrogram depend on the specific application and the characteristics of the data one is working with. The dependence of spectrogram on various factors is given in (1).

$$\text{Spectogram Length} = 1 + \text{CEIL} \frac{(SL-W)}{(W-O)} \quad (1)$$

Where, 'SL' or signal length is the length of the signal to analyse; 'W' or window length is the length of the window used to calculate spectrogram; 'O' or overlap is the amount of overlap between windows; and NFFT are the number of FFT points used in spectrogram calculation. Similarly, the time resolution (Δt) represents the width of each time bin in the spectrogram. Given sampling frequency (Fs), it can be calculated using the window size and the sampling frequency as follows:

$$\Delta t = W / Fs \quad (2)$$

The time requirement refers to the total time (T) needed to extract the spectrogram out of the signal. To calculate this, we need to consider (3):

$$T = \text{Number of samples} / Fs \quad (3)$$

In our case, we generated each frame with frame length (FL) of 1024 samples, while 8 symbols per samples (SPS) were used. So, number of symbols per frame were eventually 1024/8 = 128 symbols. Knowing the sampling frequency (Fs) of 200,000 Hz; each symbol duration was calculated as 1/(200,000 x 8) or 0.625 μsec.

*B RESOLUTION TRANSFORMATION*
The choice of window size and NFFT depends on the specific application and the desired frequency or time resolution of the spectrogram. While generating spectrograms for AMC purposes, [8] and [11] have not mentioned the exact parameters, while they used Median and Gaussian filters respectively to improve the feature extraction process. Similarly, [12] uses image enhancement techniques to improve the results. The size of spectrograms generated in [9] times around 35 ms. Other works on spectrogram relating speech [50] and motion recognition [51] records the images with length 1 sec and FFT size around 20,000 respectively. Therefore, considering it may be useful to use larger FFT sizes to achieve finer frequency resolution, along with larger window sizes to achieve better time resolution, spectrograms with good resolution having 8192 NFFT were generated from the I/Q samples of the dataset RML2016.10A. The process was repeated in combination with different window sizes and overlapping factors. Figure 7 depicts the spectrogram with the finest resolution attained, encompassing a Kaiser window of size 4096 samples, with an overlap of 3584 (87.5%). To calculate the time 'T' corresponding to 8192 samples through (3), we find that at a sampling frequency of 200,000 Hz, it corresponds to approximately 40 ms. This, accompanied with high NFFT value makes these spectrograms computationally complex and time-consuming to generate for large dataset samples and therefore not suitable for stand-alone applications. Moreover, these spectrograms suffer from loss of temporal behaviour of signal due to large window settings.

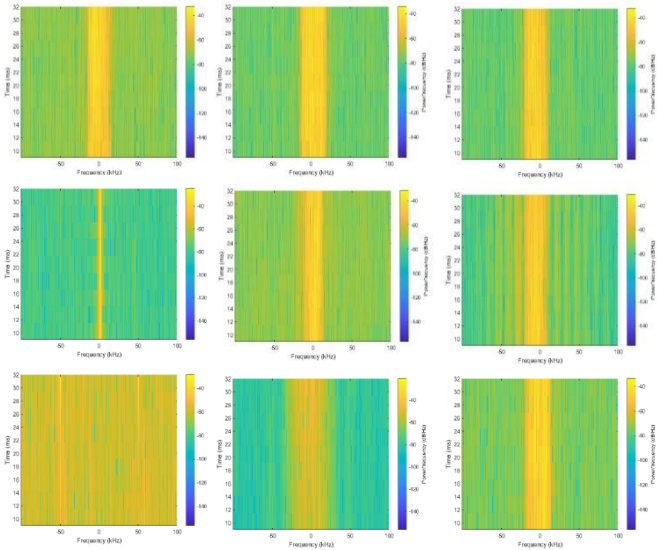

**Figure 7.** Spectrogram images of (a) 8PSK, (b) 16QAM, (c) 64QAM, (d) B-FM, (e) BPSK, (f) C-PSK, (g) DSB-AM, (h) GFSK, (j) PAM4 modulation types with NFFT size 8192 and window size 4096 with 50% overlap, resulting in spectrogram length of 4096 (1).

*C IMPROVING SPECTROGRAM CLASSIFICATION*

In [52], the time-frequency spectrograms of emitter signals was computed emphasizing on the significant features of the spectrogram. Similarly, [53] explored resolution transformation of spectrograms for harmonic protection of meshed grids. In various researches, the following guidelines were followed for improving spectrogram classification:

1) INCREASING THE AMOUNT OF TRAINING DATA

More training data can help the model learn better features and improve its accuracy. If the amount of training data is limited, data augmentation techniques may be used to generate more data [54]. Therefore, we generated more samples and extracted around 22,000 spectrograms for each SNR.

2) SELECTION OF LOSS FUNCTION

In order to improve robustness against outliers and noise, we selected a less-sensitive loss function.

3) OPTIMIZING THE HYPERPARAMETERS

Hyperparameters are the parameters that are set before training the model to help it learn better [2]. In our setup, we optimized the learning rate, batch size, and dropout for said purpose.

4) USING REGULARIZATION TECHNIQUES

To prevent overfitting and improve the model's generalization performance, dropout and L2 regularization techniques were used.

5) FINE-TUNE THE MODEL

During our experiment, we monitored models' performance on validation along with loss values to ascertain fine-tuning of models on our dataset.

In order to utilize unfiltered spectrograms for feature extraction, above mentioned guidelines were followed. Moreover, resolution transformation of generated spectrogram was carried out in light of the mathematical calculations (1-3) with aim to reduce the computational requirement as well as the image generation time. For this, different frame sizes, window lengths, overlap values and NFFT were utilized. Reduction in these parameters were done step-wise and tested for feature extraction, in order to ascertain their suitability for training the DL models. The challenge of the FFT (Fast Fourier Transform), a computationally demanding technique, is directly correlated with the number of points employed in the transformation. The FFT operation's computational complexity is typically represented as $O(N \log N)$, where N is the number of points.

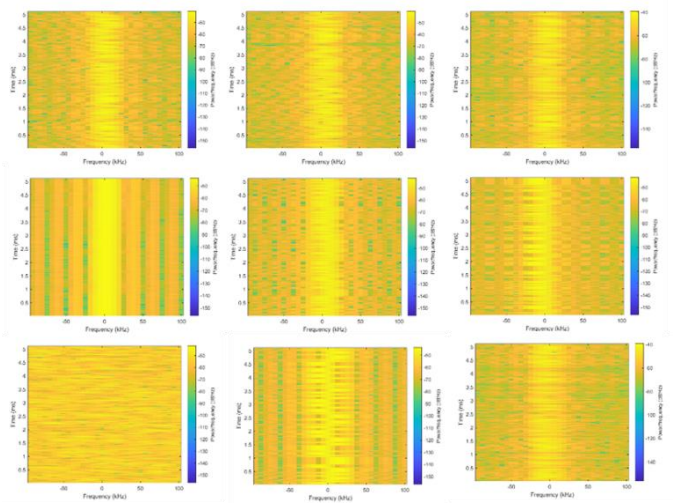

**Figure 8.** Transformed spectrograms of (a) 8PSK, (b) 16QAM, (c) 64QAM, (d) B-FM, (e) BPSK, (f) C-PSK, (g) DSB-AM, (h) GFSK, (j) PAM4 modulation types with NFFT size 32 and window size 8, with 50% overlap, resulting in spectrogram length of only 256 (1).

The amount of computations needed for the FFT process was greatly reduced by lowering the NFFT from 8192 to 32. Eventually, window size of 8 was selected for generating the spectrogram images, each from 1024 samples of I/Q data. The required time for 1024 samples was calculated (3).

We calculated that 1024 samples at given sampling frequency corresponds to approximately 5.12 milliseconds, thereby making it 8x more time efficient spectrogram if compared to the one previously generated. Sample of these transformed spectrograms is shown in figure 8. Moreover, while comparing the computational requirements of previous and the transformed spectrograms, 99.61% of computational workload reduction was calculated using NFFT of 32 instead of 8192.

*D CREATING LABELLED DATASET FOR TRAINING*

In order to verify the effectiveness of the proposed algorithm, we utilized the transformed images. For this purpose, a dataset was generated, pre-processed and resized for selected DL models mentioned in table 9 with multiple training options for fine tuning to get the results. Size of dataset used was 1940 samples of each modulation type (21,340 in total for 11 modulation types) and testing was carried out at multiples SNRs from 5 dB to 30 dB, with a step of 5 dB on selected DL models.

TABLE IX
PARAMETRIC COMPARISON OF DL MODELS UTILIZED

| DL Model | Ref | Input Size | Top-1 Acc. | Top-5 Acc. | FLOPS (Bil) | Para (Mil) |
|---|---|---|---|---|---|---|
| Dense Net201 | [55] | 224x224 | 77.6% | 93.8% | 6.4 | 20 |
| ResNet-50 | [56] | 224x224 | 76.1% | 92.9% | 4.1 | 25.6 |
| InceptionV3 | [57] | 299x299 | 78.8% | 94.4% | 5.7 | 23.9 |
| IncResNetV2 | [58] | 299x299 | 80.2% | 95.2% | 11.2 | 55.9 |
| SqueezeNet | [59] | 224x224 | 57.5% | 80.3% | 0.829 | 0.72 |
| VGG16 | [60] | 224x224 | 71.5% | 90.2% | 15.3 | 138.3 |

The term *input size* relates to the size of the images that the model is intended to analyse. The number of mathematical operations needed to perform inference on an input image is measured in *FLOPS* (floating-point operations per second), which can be used to evaluate the computational complexity of various models. The capacity to deploy the model on various devices might be impacted by the model's parameters. *Top-1* and *top-5* accuracy, respectively, refer to the model's precision in identifying the proper class for the top-1 and top-5 most likely predictions.

## VII. EXPERIMENTAL RESULTS AND EVALUATION

During the experiments, we compared the classification accuracy of our dataset against existing CNN models. The recognition accuracy of all said models under different SNR were evaluated. SNR refers to the ratio of signal to noise in an electronic equipment. In general, the higher the signal to noise ratio, the lower the amount of noise that has been mixed with the signal, or vice versa. In addition, we also analysed the recognition rate of each model with respect to different modulation schemes. The basic properties of these CNN-based models were analysed before undertaking the simulations. Table IX shows their comparison in terms of vital characteristics while figure 9 shows the results obtained when the multiple CNN models were trained to classify all 11 modulation types. The highest classification performance was obtained with InceptionResnetV2 and VGG-16. As shown that except QAM signals inter-confusion due to phase and frequency off-set, all other spectrograms were measured with good accuracy. The trained model was tested with random samples and the results obtained are depicted in figure 10.

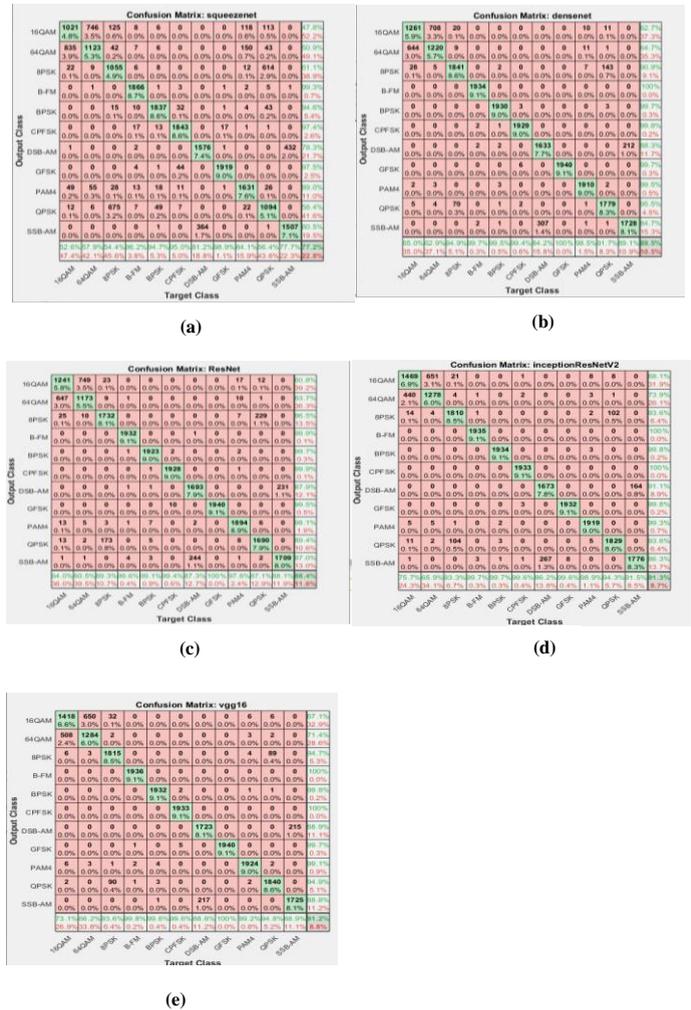

Figure 9. Confusion matrix depicting the accuracy of deep learning models (a) SqueezeNet. (b) DenseNet-201. (c) Resnet-50. (d) IncResnt-V2. (e) VGG-16 - against sample size of 1940 per class.

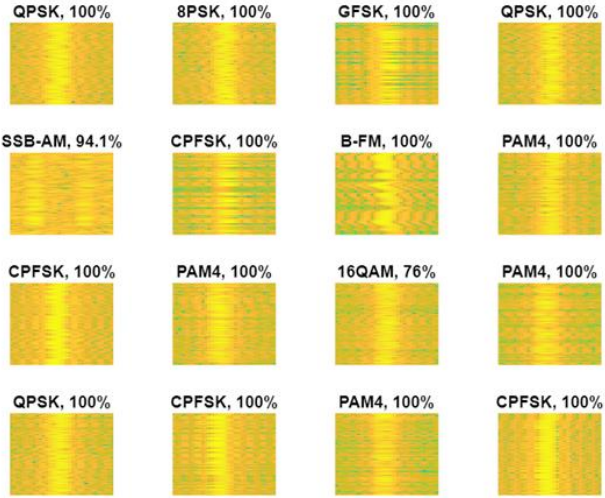

**Figure 10.** Random sample classification accuracy while testing a spectrogram-trained Resnet-50.

The deep neural network communication signal modulation classifier was implemented on CPU and GPU hardware platform, and all the training and testing tasks have been completed. Experiments show that all models used for experiment have good performance on both the training set and the test set, and can accurately identify the modulation mode of the communication signal at different SNRs from 05-30 dB, which shows the robustness of the spectrogram data. According to the simulations, SqueezeNet achieves the lowest accuracy of 77.2%, whereas other models achieve around 90%. The best results were observed for IncResnetV2 with an accuracy of 91.1%. Figure 11 (a) shows the comparative performance of different models w.r.t. modulation type, while figure 11 (b) depicts the performance under various SNR conditions.

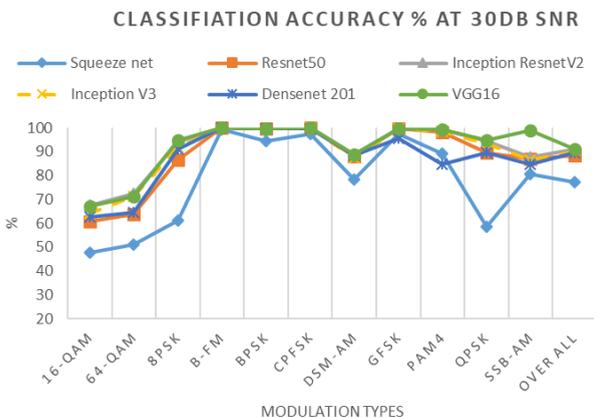

**Figure 11. (a)** The comparative performance of different DL models w.r.t. modulation type.

Inter-class confusion exists in all the results among 16-QAM and 64-QAM. The reason for this confusion may be analysed through signal visualization. It can be observed from Figure 7 and 8 that the spectrogram of each modulation type has unique features, but those of 16-QAM and 64-QAM are very similar in presence of noise, leading to low classification rates. In addition, AM-DSB and AM-SSB also possess some similarities thereby resulting in error while classifying these two modulation schemes. The model was also trained with 10 modulation types, excluding 64QAM in order to validate the improved classification accuracy, the result after said adjustment were increased till 96% where all modulation types had *accuracy close to 100%,* except 8PSK and DSB-AM which had accuracy around 89.4% and 81.7% respectively. The confusion matrix and training progress for this is also depicted in figure 12.

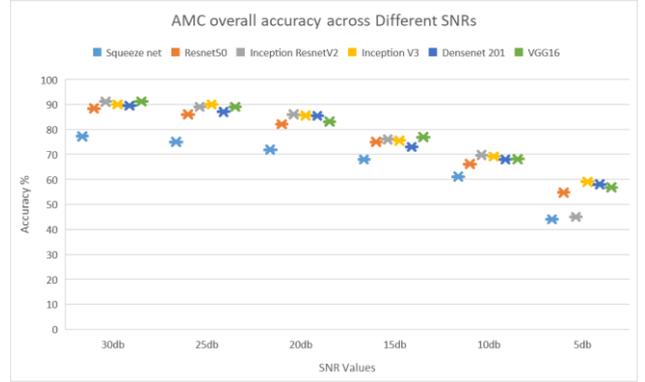

**Figure 11. (b)** Overall classification accuracy under various SNR conditions.

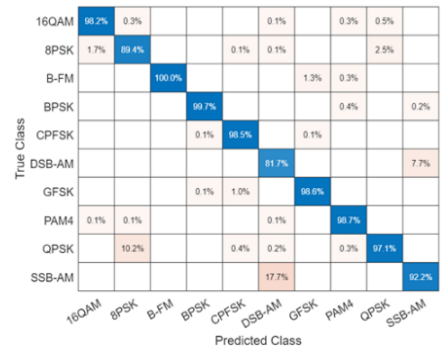

**Figure 12.** The confusion matrix for 10 modulation type with 96% accuracy when trained with Resnet50.

*A TRAINING PARAMETERS AND SETTINGS*

For our AMC models to perform the classification tasks; hidden nodes in the layers, network depth, activation functions, and other parameters are all established. In our design, the network is trained with only 30 epochs and network, with forced stopping of training with suitable validation patience and output was sought for best validation loss. The Adam optimizer was used to acquire the best-learned weights. The mini-batch size was kept 64 images for each training iterations due to GPU performance limitation (NVIDIA 3060). The initial learning rate for training is 0.0001. The learning rate was scheduled piecewise, the software reduced it every 5 epochs by a factor of 0.1. The validation frequency was kept as 30 so that the network may be validated after 30 iterations during the training process, while L2 regularization was kept at 0.0001. An Early Stopping was applied in order to avoid overfitting and the network output was saved for best validation loss.

## VIII. CONCLUSIONS AND FUTURE DEVELOPMENTS

Two main contributions of this work are to demonstrate DL based AMC and to show the feature extraction capability of various CNN models from transformed spectrograms that have low computational requirements. For this, a framework for end-to-end learning from raw spectrograms was proposed. The results support the conclusion that the transfer learning of spectrogram-based approach is a suitable solution for AMC under various SNRs. They also prove that it is possible to develop a real-time modulation classifier for the given modulation schemes. Since our experiment is based on an end-to-end Deep Learning model, applied directly to raw spectrograms without a manual feature extraction procedure; these unfiltered images enable us to readily pre-process and train our model; thereby enhancing the practicability of this model under various environments. The results also showed that, for the wireless communication domain, it is crucial to investigate several wireless data representations in order to select the one that not only exhibits discriminative properties for the signals being categorised but also provides a resource-efficient solution. Two active wireless signal identification research issues can be addressed using the presented approach: (i) modulation recognition, which is important for applications requiring dynamic spectrum access; and (ii) wireless interference identification, which is crucial for effective interference mitigation techniques in unrestricted bands. By implementing the suggested methodology, DL / DSP practitioners and wireless engineers may get improved understandings of the ideal data representations for other DL-based classifications in future. Furthermore, researchers may focus on identifying diverse types of digital signals at low SNRs along with addressing the challenges associated with classifying modulation under varying noise conditions. Utilizing spectrum data effectively can address problems relating to EM interferences, ineffective spectrum usage, and regulation in future.